\documentstyle[aps,prl,epsfig,floats]{revtex}

\begin{document}

\twocolumn[\hsize\textwidth\columnwidth\hsize\csname@twocolumnfalse\endcsname

\title{Controlling Spin Exchange Interactions of Ultracold Atoms in Optical
Lattices}

\author{L.-M. Duan$^{1}$, E. Demler$^{2}$, and M. D. Lukin$^{2}$}
\address{$^1$ Institute for quantum information, California Institute of
Technology, mc 107-81, Pasadena, CA 91125}
\address{$^2$ Physics Department, Harvard University, Cambridge, Massachusetts
02138}
\maketitle

\begin{abstract}
We describe a general technique that allows to induce and control
strong interaction between spin states of neighboring atoms in an
optical lattice.  We show that the properties of spin exchange
interactions, such as magnitude, sign, and anisotropy can be
designed by adjusting the optical potentials. We illustrate how
this technique can be used to efficiently ``engineer'' quantum
spin systems with desired properties, for specific examples
ranging from scalable quantum computation to probing a model with
complex topological order that supports exotic
anyonic excitations.

PACS numbers:03.75.Fi,03.67.-a,71.10.-w,42.50.-p
\end{abstract}
 ]

Recent observations of the superfluid to Mott insulator transition in a
system of ultracold atoms in an optical lattice open fascinating prospects
for studying many body phenomena associated with strongly correlated systems
in a highly controllable environment \cite{3,4,2,5}.
For instance, the recent studies have shown that, with spinor bosonic or
fermionic atoms in optical lattices, it may be possible to observe complex
quantum phase transitions \cite{6,6'}, to probe novel superfluidity
mechanisms \cite{7,7'}, or to demonstrate the spin-charge separation
predicted from the Luttinger liquid model \cite{8}.

This Letter describes a general technique to control many-body spin
Hamiltonians using ultra-cold atoms. Specifically, we show that when
two-state bosonic or fermionic atoms are confined in an optical lattice, the
interaction between spins of the particles can be controlled by adjusting
the intensity, frequency, and polarization of the trapping light. The
essential idea is to induce and control virtual spin-dependent tunneling
between neighboring atoms in the lattice that results in a controllable
Heisenberg exchange interaction. By combining this simple experimental
technique with the design of the lattice geometry, it is possible to
``engineer'' many interesting spin Hamiltonians corresponding to strongly
correlated systems.

Such techniques are of particular significance since quantum magnetic
interactions are central to understanding complex orders and correlations
\cite{auerbach}. We illustrate this with several examples: (i) we show that
one of the generated Hamiltonians provides us an easy way to realize the
so-called cluster states in two or three dimensions \cite{9}, which are
useful for an implementation of scalable quantum computation with neutral
atoms; (ii) we show that the realized Hamiltonian has a rich phase diagram,
opening up the possibility to observe various quantum magnetic phase
transitions in a controllable way; (iii) finally, we show how to implement
an exactly-solvable spin Hamiltonian recently proposed by Kitaev \cite{10},
which supports abelian and non-abelian anyonic excitations with exotic
fractional statistics. Abelian anyons could also exist in a fast rotating
condensate \cite{11}.

We consider an ensemble of ultracold bosonic or fermionic atoms confined in
an optical lattice formed by several standing wave laser beams. We are
interested in the Mott insulator regime, and the atomic density of roughly
one atom per lattice site. Each atom is assumed to have two relevant
internal states, which are denoted with the effective spin index $\sigma
=\uparrow ,\downarrow $, respectively. We assume that the atoms with spins $%
\sigma =\uparrow ,\downarrow $ are trapped by independent standing wave
laser beams through polarization (or frequency) selection. Each laser beam
creates a periodic potential $V_{\mu \sigma }sin^{2}(\vec{k}_\mu\cdot \vec{r}%
)$ in a certain direction $\mu $, where $\vec{k}_\mu$ is the wavevector of
light. For sufficiently strong periodic potential and low temperatures, the
atoms will be confined to the lowest Bloch band as has been confirmed from
experiments \cite{3}, and the low energy Hamiltonian is then given by
\begin{eqnarray}
H &=&-\sum_{\langle ij\rangle \sigma }\left( t_{\mu \sigma }a_{i\sigma
}^{\dagger }a_{j\sigma }+H.c.\right)  \nonumber \\
&&+\frac{1}{2}\sum_{i,\sigma }U_{\sigma }n_{i\sigma }\left( n_{i\sigma
}-1\right) +U_{\uparrow \downarrow }\sum_{i}n_{i\uparrow }n_{i\downarrow },
\label{Hamiltonian1}
\end{eqnarray}
Here $\left\langle i,j\right\rangle $ denotes the near neighbor sites in the
direction ${\mu }$, $a_{i\sigma }$ are bosonic (or fermionic) annihilation
operators respectively for bosonic (or fermionic) atoms of spin $\sigma $
localized on site $i$, and $n_{i\sigma }=a_{i_{\sigma }}^{\dagger
}a_{i_{\sigma }}$.

For the cubic lattice ($\mu =x$,$y$,$z$) and using a harmonic approximation
around the minima of the potential \cite{2}, the spin-dependent tunneling
energies and the on-site interaction energies are given by $t_{\mu \sigma
}\approx \left( 4/\sqrt{\pi }\right) E_{R}^{1/4}\left( V_{\mu \sigma
}\right) ^{3/4}\exp [-2(V_{\mu \sigma }/E_{R})^{1/2}]$, $U_{\uparrow
\downarrow }\approx (8/\pi )^{1/2}(ka_{s\uparrow \downarrow })(E_{R}%
\overline{V}_{1\uparrow \downarrow }\overline{V}_{2\uparrow \downarrow }%
\overline{V}_{3\uparrow \downarrow })^{1/4}$. Here $\overline{V}_{\mu
\uparrow \downarrow }=4V_{\mu \uparrow }V_{\mu \downarrow }/(V_{\mu \uparrow
}^{1/2}+V_{\mu \downarrow }^{1/2})^{2}$ is the spin average potential in
each direction, $E_{R}=\hbar ^{2}k^{2}/2m$ is the atomic recoil energy, and $%
a_{s\uparrow \downarrow }$ is the scattering length between the atoms of
different spins. For bosonic atoms $U_{\sigma }\approx (8/\pi )^{1/2}\left(
ka_{s\sigma }\right) \left( E_{R}V_{1\sigma }V_{2\sigma }V_{3\sigma }\right)
^{1/4}$ ($a_{s\sigma }$ are the corresponding scattering lengths). For
fermionic atoms, $U_{\sigma }$ is on the order of Bloch band separation $%
\sim 2\sqrt{V_{\mu \sigma }E_{R}}$, which is typically much larger than $%
U_{\uparrow \downarrow }$ and can be taken to be infinite. In writing Eq.
(1), we have neglected overall energy shifts $\sum_{i\mu }\left( \sqrt{%
E_{R}V_{\mu \uparrow }}-\sqrt{E_{R}V_{\mu \downarrow }}\right) \left(
n_{i\uparrow }-n_{i\downarrow }\right) /2$, which can be easily compensated
by a homogeneous external magnetic field applied in the $z$ direction.

From the above expressions, we observe that $t_{\mu \sigma }$ depend
sensitively (exponentially) upon the ratios $V_{\mu \sigma }/E_{R}$ while $%
U_{\uparrow \downarrow }$ and $U_{\sigma }$ exhibit only weak dependence. So
we can easily introduce spin-dependent tunneling $t_{\mu \sigma }$ by
varying the potential depth $V_{\mu \uparrow }$ and $V_{\mu \downarrow }$
with control of the intensity of the trapping laser. We now show that this
simple experimental method provides us a powerful tool to engineer many-body
Hamiltonians. We are interested in the regime where $t_{\mu \sigma }\ll
U_{\sigma },U_{\uparrow \downarrow }$ and $\left\langle n_{i\uparrow
}\right\rangle +\left\langle n_{i\downarrow }\right\rangle \simeq 1$, which
corresponds to an insulating phase. In this regime, the terms proportional
to tunneling $t_{\mu \sigma }$ can be considered via perturbation theory. We
use a simple generalization of the Schriffer-Wolf transformation \cite
{hewson} (see another method in \cite{7'}), and to the leading order in $%
t_{\mu \sigma }/U_{\uparrow \downarrow }$, Eq. (1) is equivalent to the
following effective Hamiltonian
\begin{equation}
H=\sum_{\left\langle i,j\right\rangle }\left[ \lambda _{\mu z}\sigma
_{i}^{z}\sigma _{j}^{z}\pm \lambda _{\mu \perp }\left( \sigma _{i}^{x}\sigma
_{j}^{x}+\sigma _{i}^{y}\sigma _{j}^{y}\right) \right] ,
\label{Hamiltonian2}
\end{equation}
Here $\sigma _{i}^{z}=n_{i\uparrow }-n_{i\downarrow }$, $\sigma
_{i}^{x}=a_{i\uparrow }^{\dagger }a_{i\downarrow }+a_{i\downarrow }^{\dagger
}a_{i\uparrow }$, and $\sigma _{i}^{y}=-i\left( a_{i\uparrow }^{\dagger
}a_{i\downarrow }-a_{i\downarrow }^{\dagger }a_{i\uparrow }\right) $ are the
usual spin operators. The $+$ and $-$ signs before $\lambda _{\mu \perp }$
in Eq. (4) correspond respectively to the cases of fermionic and bosonic
atoms. The parameters $\lambda _{\mu z}$ and $\lambda _{\mu \perp }$ for the
bosonic atoms are given by
\begin{equation}
\lambda _{\mu z}=\frac{t_{\mu \uparrow }^{2}+t_{\mu \downarrow }^{2}}{%
2U_{\uparrow \downarrow }}-\frac{t_{\mu \uparrow }^{2}}{U_{\uparrow }}-\frac{%
t_{\mu \downarrow }^{2}}{U_{\downarrow }},\;\;\lambda _{\mu \perp }=\frac{%
t_{\mu \uparrow }t_{\mu \downarrow }}{U_{\uparrow \downarrow }}.
\label{lambda}
\end{equation}
For fermionic atoms the expression for $\lambda _{\perp }$ is the same as in
(\ref{lambda}), but in the expression for $\lambda _{z}$ the last two terms
vanish since $U_{\sigma }\gg U_{\uparrow \downarrow }$. In writing Eq. (2),
we neglected the term $\sum_{i\mu }4\left( t_{\mu \uparrow }^{2}/U_{\uparrow
}-t_{\mu \downarrow }^{2}/U_{\downarrow }\right) \sigma _{i}^{z}$, which can
be easily compensated by an applied external magnetic field.

The Hamiltonian (2) represents the well-known anisotropic Heisenberg model ($%
XXZ$ model), which arises in the context of various condensed matter systems
\cite{auerbach}. However, the approach involving ultracold atoms has a
unique advantage in that the parameters $\lambda _{\mu z}$ and $\lambda
_{\mu \perp }$ can be easily controlled by adjusting the intensity of the
trapping laser beams. They can also be changed within a broad range by
tuning the ratio between the scattering lengths $a_{s\uparrow \downarrow }$
and $a_{s\sigma }$ $\left( \sigma =\uparrow ,\downarrow \right) $ by
adjusting an external magnetic field through Feshbach resonance \cite{15}.
Therefore, even with bosonic atoms alone, it is possible to realize the
entire class of Hamiltonians in the general form (\ref{Hamiltonian2}) with
an arbitrary ratio $\lambda _{\mu z}/\lambda _{\mu \perp }$. This is
important since bosonic atoms are generally easier to cool. In Figure 1a we
show the phase diagram of the Hamiltonian (\ref{Hamiltonian2}) on a
bipartite lattice as a function of $\beta _{t}=t_{\uparrow }/t_{\downarrow
}+t_{\downarrow }/t_{\uparrow }$ and $U_{\uparrow \downarrow }/U_{\sigma }$
\cite{16a} for the case when $U_{\uparrow }=U_{\downarrow }$ and $t_{\mu
\sigma }$ independent of the spatial direction $\mu $. Certain lines on this
phase diagram correspond to well known spin systems: when $U_{\uparrow
\downarrow }/U_{\sigma }=1/2$ we have an $XY$ model; when $\beta _{t}=\infty
$ ($t_{\uparrow }$ or $t_{\downarrow }$ is zero) we have an Ising model; for
$\beta _{t}=\pm (1/2-U_{\uparrow \downarrow }/U_{\sigma })^{-1}$ we have an $%
SU(2)$ symmetric antiferromagnetic or ferromagnetic systems, respectively.

Before proceeding, we estimate the typical energy scales and discuss the
influence of imperfections and noise. For Rb atoms with a lattice constant $%
\pi /\left| \vec{k}\right| \sim 426$nm, the typical tunnelling rate $t/\hbar
$ can be chosen from zero to a few kHz \cite{3}. The on-site interaction $%
U/\hbar $ corresponds to a few kHz at zero magnetic field, but can be much
larger near the Feshbach resonance. The energy scale for magnetic
interaction is about $t^{2}/\hbar U\sim 0.1$kHz (corresponding to a time
scale of $10$ms) with a conservative choice of $U\sim 2$kHz and $\left(
t/U\right) ^{2}\sim 1/20$. These energy scales are clearly compatible with
current experiments \cite{3}. We further note that the present system should
be quite robust to realistic noise and imperfections. First of all, the next
order correction to the Hamiltonian (2) is proportional to $\left(
t/U\right) ^{2}$, which is small in the Mott regime. Second, since the atoms
only virtually tunnel to the neighboring sites with a small probability $%
\left( t/U\right) ^{2}$, the dephasing rate and the inelastic decay rate are
significantly reduced compared with the cold collision scheme \cite{16,19}.
Finally, the spontaneous emission noise rate can be made very small by using
a blue-detuned optical lattice or by increasing the detuning. In a
blue-detuned lattice, even with a moderate detuning $\Delta \sim 5$GHz, the
effective spontaneous emission rate is estimated to be of the order of Hz,
which is significantly smaller than $t^{2}/\left( \hbar U\right) $.

We now illustrate the ability to engineer many-body spin Hamiltonians with
specific examples. For the first example, we set $V_{\mu \downarrow }/V_{\mu
\uparrow }\gg 1$, so that $t_{\mu \downarrow }$ becomes negligible while $%
t_{\mu \uparrow }$ remains finite. In this case, the Hamiltonian (2) reduces
to the Ising model $H=\sum_{\left\langle i,j\right\rangle }\lambda _{\mu
z}\sigma _{i}^{z}\sigma _{j}^{z}$, with $\lambda _{\mu z}=t_{\mu \uparrow
}^{2}/\left( 0.5/U_{\uparrow \downarrow }-1/U_{\uparrow }\right) $. Though
this Hamiltonian has quite trivial properties for its ground states and
excitations, its realization in optical lattices can be very useful for a
dynamical generation of the so-called cluster states \cite{9}. Specifically,
we note that this Ising interaction can be easily turned on and off by
adjusting the potential depth $V_{\mu \uparrow }$. If we first prepare each
atom in the lattice into the superposition state $\left( \left| \uparrow
\right\rangle +\left| \downarrow \right\rangle \right) /\sqrt{2}$, and then
lower $V_{\mu \uparrow }$ for a time $T$ with $\lambda _{\mu z}T=\pi /4$ mod
$\pi /2$, the final state is a cluster state with its dimension determined
by the dimension of the lattice \cite{9}. The $d$-dimensional ($d\geq 2$)
cluster states have important applications for implementation of scalable
quantum computation with neutral atoms: after its preparation, one can
implement universal quantum computation simply via a series of single-bit
measurements only \cite{9}. The use of such cluster states can significantly
alleviate the stringent requirements on separate addressing of the
neighboring atoms in the proposed quantum computation schemes \cite{16,17}.
Although the present approach is somewhat slower that the cold collision
scheme \cite{16}, it allows to take advantage of its simplicity and the
reduced dephasing rate.

As our second example, we explore the rich phase diagram of the Hamiltonian (%
\ref{Hamiltonian2}) in the presence of magnetic fields. For simplicity, we
assume a bipartite lattice and identical spin exchange constants for all
links. Figure 1b shows the mean-field phase diagram for bosonic particles in
the presence of a longitudinal field $h_z$. This diagram was obtained by
comparing energies of the variational wavefunctions of two kinds: (i) the
Neel state in the $z$ direction $\left\langle \vec{\sigma}_{i}\right\rangle
=(-1)^{i}\vec{e}_{z}$; (ii) canted phase with ferromagnetic order in the $xy$
plane and finite polarization in the $z$ direction $\left\langle \vec{\sigma}%
_{i}\right\rangle = \vec{e}_{x}cos\theta +\vec{e}_{z} sin\theta$. Here, $%
\theta $ is a variational parameter, and $\vec{e}_{z,x}$ are unit vectors in
the directions $z,x$. Transition between the $z$-Neel and the canted phases
is a first order spin-flop transition \cite{fisher} at $h_z=Z (\lambda_z^2 -
\lambda_{\perp}^2)^{1/2}$ ($Z$ is the number of neighboring atoms of each
lattice site), and transition between the $xy$-Neel phase and the $z$
polarized phase is a second order transition of the $XY$ type at $h_z = Z
(\lambda_z + \lambda_{\perp})$. In the absence of transverse magnetic field
one can use the existence of two sub-lattices to change the sign of $%
\lambda_{\perp}$ using the transformation $\sigma^{x,y}_i \rightarrow (-)^i
\sigma^{x,y}_i$. Hence, fermionic atoms in the longitudinal magnetic field
have the same phase diagram as shown in Figure 1b, except that their canted
phase has transverse Neel rather than transverse ferromagnetic order.
Results of a similar mean-field analysis of the Hamiltonian (\ref
{Hamiltonian2}) for bosonic atoms with a transverse magnetic field $h_x$ are
shown in Fig. 1c. For fermionic atoms in a transverse field, there is one
more phase with a Neel order along $y$ direction.

\begin{figure}[tbp]
\epsfig{file=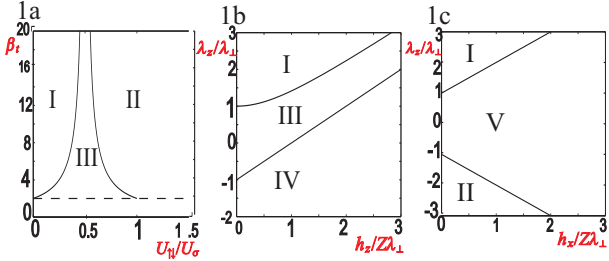,width=8cm}
 \caption{Phase diagrams of the Hamiltonians (\ref{Hamiltonian2}) for bosonic and fermionic
atoms: (a) at zero magnetic field, (b) with a longitudinal field $h_{z}$, and (c) (for bosons only) with a transverse field
$h_{x}$. Each phase is characterized by the following order parameter: I - $z$-Neel order; II - $z$-ferromagnetic order; III -
$xy$-Neel order for fermionic atoms and $xy$-ferromagnetic order for bosonic atoms; IV and V - spin polarization in the direction
of applied field, $z$ and $x$ respectively.}
\end{figure}

The third example involves the anisotropic spin model on a 2-D hexagonal
lattice proposed recently by Kitaev \cite{10}. In this model interactions
between nearest neighbors are of the XX, YY or ZZ type, depending on the
direction of the link
\begin{equation}
H=\sum_{\nu =x,y,z;{\left\langle i,j\right\rangle \in D_{\nu }}}\lambda
_{\nu }\sigma _{i}^{\nu }\sigma _{j}^{\nu },  \label{kitaev}
\end{equation}
where the symbol $\left\langle i,j\right\rangle \in D_{\nu }$ denotes the
neighboring atoms in the $D_{\nu }$ $\left( \nu =x,y,z\right) $ direction
(see Fig. 2b).

\begin{figure}[tbp]
\epsfig{file=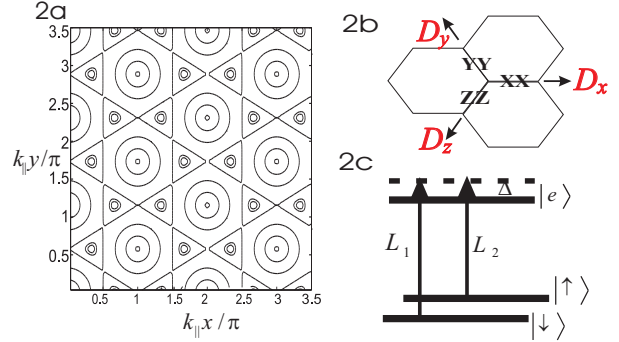,width=8cm} \caption{2a. The contours with the three potentials in the form of Eq. (5).
The minima are at centers of the triangles when $\protect\varphi _{0}=%
\protect\pi /2$. (2b). The illustration of the model Hamiltonian (4). (2c). The schematic atomic level structure and the laser
configuration to induce spin-dependent tunneling. }
\end{figure}

To implement this model using ultracold atoms, we first raise the potential
barriers along the vertical direction Z \ in the 3-dimensional (3D) optical
lattice so that the tunneling and the spin exchange interactions in Z
direction are completely suppressed \cite{8,3}. In this way, we get an
effective 2D configuration with a set of independent identical 2D lattice in
the X-Y plane. We then apply in the X-Y plane three trapping potentials
(identical for all spin states) of the forms
\begin{equation}
V_{j}\left( x,y\right) =V_{0}\sin ^{2}\left[ k_{\parallel }\left( x\cos
\theta _{j}+y\sin \theta _{j}\right) +\varphi _{0}\right] ,
\end{equation}
where $j=1,2,3$, and $\theta _{1}=\pi /6$, $\theta _{2}=\pi /2$, $\theta
_{3}=-\pi /6$. Each of the potentials is formed by two blue-detuned
interfering traveling laser beams above the X-Y plane with an angle $\varphi
_{\parallel }=2\arcsin \left( 1/\sqrt{3}\right) $, so that the wave vector $%
k_{\parallel }$ projected onto the X-Y plane has the value $k_{\parallel
}=k\sin \left( \varphi _{\parallel }/2\right) =k/\sqrt{3}$. We choose the
relative phase $\varphi _{0}=\pi /2$ in Eq. (5) so that the maxima of the
three potentials overlap. In this case, the atoms are trapped at the minima
of the potentials, which form a hexagonal lattice as shown by the centers of
the triangles in Fig. 2a. We assume that there is one atom per each lattice
site, and this atom interacts with the three neighbors in different
directions through virtual tunneling with a potential barrier given by $%
V_{0}/4$.

In such a hexagonal lattice, we wish to engineer anisotropic Heisenberg
exchange for each tunneling direction (denoted by $D_{x}$, $D_{y}$, and $%
D_{z}$, respectively). To this end, we apply three blue-detuned
standing-wave laser beams in the X-Y plane along these tunneling directions:
\begin{equation}
V_{\nu \sigma }\left( x,y\right) =V_{\nu \sigma }\sin ^{2}\left[ k\left(
x\cos \theta _{\nu }^{\prime }+y\sin \theta _{\nu }^{\prime }\right) \right]
,
\end{equation}
where $\nu =x,y,z$, and $\theta _{x}^{\prime }=-\pi /3$, $\theta
_{y}^{\prime }=\pi $, $\theta _{z}^{\prime }=\pi /3$. In general, we require
that the potential depth $V_{\nu \sigma }$ depend on the atomic spin state
as
\begin{equation}
V_{\nu \sigma }=V_{\nu +}\left| +\right\rangle _{\nu }\left\langle +\right|
+V_{\nu -}\left| -\right\rangle _{\nu }\left\langle -\right| ,\text{ }\left(
\nu =x,y,z\right) ,
\end{equation}
where $\left| +\right\rangle _{\nu }$ ($\left| -\right\rangle _{\nu }$) is
the eigenstate of the corresponding Pauli operator $\sigma ^{\nu }$ with the
eigenvalue $+1$ ($-1$).

The spin-dependent potentials in the form of Eqs. (6,7) can be realized, for
instance, with the specific atomic level configuration shown in Fig. 2c.
Here, $\sigma =\uparrow ,\downarrow $ denote two hyperfine levels of the
atom with different energies. They are coupled to the common excited level $%
\left| e\right\rangle $ with a blue detuning $\Delta $, respectively through
the laser beams $L_{1}$ and $L_{2}$ with frequencies matching the
corresponding transitions. The quantization axis is chosen to be
perpendicular to the X-Y plane, and the phase-locked laser beams $L_{1}$ and
$L_{2}$ are both polarized along this direction. In the tunneling direction $%
D_{z}$, we only apply the $L_{1}$ laser beam, which induces the potential $%
V_{z\sigma }\left( x,y\right) $ with the desired form (7) of its depth $%
V_{z\sigma }$. In the tunneling direction $D_{x}$ or $D_{y}$, we apply both
lasers $L_{1}$ and $L_{2}$, but with different relative phases, which
realize the desired potential depth $V_{x\sigma }$ or $V_{y\sigma }$ of the
form (7) in the corresponding direction.

The potentials (6,7) do not have influence on the equilibrium positions of
the atoms, but they change the potential barrier between the neighboring
atoms in the $D_{\nu }$ direction from $V_{0}/4$ to $V_{\nu \sigma }^{\prime
}=V_{0}/4+V_{\nu \sigma }$. The parameters $V_{\nu +}$ and $V_{\nu -}$ in
Eq. (7) can be tuned by varying the laser intensity of $L_{1}$ and $L_{2}$
in the $D_{\nu }$ direction, and one can easily find their appropriate
values so that in the $D_{\nu }$ direction, the atom can virtually tunnel
with a rate $t_{+\nu }$ only when it is in the eigenstate $\left|
+\right\rangle _{\nu }$. Hence, it follows from Eqs.(2,3) that the effective
Hamiltonian for our system is given by Eq.(4) with $\lambda _{\nu }\approx
-t_{+\nu }^{2}/\left( 2U\right) $ for bosonic atoms with $U_{\uparrow
\downarrow }\approx U_{\uparrow }\approx U_{\downarrow }\approx U$. After
compensating effective magnetic fields, we find exactly the model described
by the Hamiltonian (\ref{kitaev}).

The model (\ref{kitaev}) is exactly solvable due to the existence of many
conserved operators, and it has been shown to possess very interesting
properties \cite{10}. In particular, it supports both abelian and
non-abelian anyonic excitations, depending on the ratios between the three
parameters $\lambda _{\nu }$. In the region where $2\lambda _{\nu }/\left(
\lambda _{x}+\lambda _{y}+\lambda _{z}\right) \leq 1,$ ($\nu =$ $x,y,z$),
the excitation spectrum of the Hamiltonian (\ref{kitaev}) is gapless, but a
gap opens when perturbation magnetic fields are applied in the $x,y,z$
directions, and the excitations in this case obey non-abelian fractional
statistics. Out of this region, except at some trivial points with $\lambda
_{x}\lambda _{y}\lambda _{z}=0$, the Hamiltonian (\ref{kitaev}) has gapped
excitations which satisfy abelian fractional statistics. Thus, the present
implementation opens up an exciting possibility to realize experimentally
the exotic abelian and non-abelian anyons.

Now we briefly discuss the techniques for probing the resulting states. To detect the quantum phase transitions in the XXZ model
with magnetic fields or in Kitaev's model, one can probe the excitation spectra via Bragg or Raman spectroscopy. In general,
different quantum phases are characterized by specific dispersion relations (for instance, in Kitaev's model, one phase is gapped
while the other is gapless). If the two probe light beams have the momentum and frequency differences which match those of the
dispersion relation in the corresponding phase, a resonant absorption of the probe light could be observed \cite{20}. The direct
observation of the fractional statistics in Kitaev's model can be based on atomic interferometry with a procedure similar to that
described in Ref. \cite{11}: one generates a pair of anyonic excitations with a spin-dependent laser focused on two lattice sites,
rotates one anyon around the other, and then brings them together for fusion which gives different results depending on the
anyonic statistics. Other methods for detecting complex quantum states of atoms have also been developed recently \cite{altman}.

In summary, we have described a general technique to engineer many-body spin Hamiltonians.

We thank J.I.Cirac, D.DiVincenzo, W. Hofstetter, A.Kitaev, and P. Zoller for helpful discussions. This work was supported by NSF
(EIA-0086038, DMR-0132874 and PHY-0134776), and by Sloan and Packard Foundations. L.M.D. also acknowledges support from the MURI
Center DAAD19-00-1-0374, the CNSF, the CAS, and the ''97.3'' project 2001CB309300.



\begin{references}
\bibitem{16a}  We note that under the conditions of Ref.\cite{3}
inhomogeneous Mott state can be created due to overall trapping potential.
This can result in a domain structure involving new, coexisting phases,
which will be discussed elsewhere.
\end{references}
\end{document}